\documentclass{llncs}

 \usepackage{amsfonts}
 \usepackage{amsmath}
 \usepackage{amssymb}
 \usepackage{graphicx}    % needed for including graphics e.g. EPS, PS
 \usepackage{algorithmic}
 \usepackage{algorithm}

\newcommand{\bbP}{\mathbb{P}}

\title{Fast Computation of the Kinship Coefficients}
\titlerunning{Kinship Coefficients}
\date{Feb. 12, 2016}

\author{B. Kirkpatrick}
\institute{Intrepid Net Computing, \email{bbkirk@intrepidnetcomputing.com}, Montana, USA}

\begin{document}
\maketitle

\begin{abstract}

For families, kinship coefficients are quantifications of the amount of genetic sharing between a pair of individuals.  These coefficients are critical for understanding the breeding habits and genetic diversity of diploid populations.  Historically, computations of the inbreeding coefficient were used to prohibit inbred marriages and prohibit breeding of some pairs of pedigree animals.  Such prohibitions foster genetic diversity and help prevent recessive Mendelian disease at a population level.

This paper gives the fastest known algorithms for computing the kinship coefficient of a set of individuals with a known pedigree.  The algorithms given here consider the possibility that the founders of the known pedigree may themselves be inbred, and they compute the appropriate inbreeding-adjusted kinship coefficients.  The exact kinship algorithm has running-time $O(n^2)$ for an $n$-individual pedigree.  The recursive-cut exact kinship algorithm has running time $O(s^2m)$ where $s$ is the number of individuals in the largest segment of the pedigree and $m$ is the number of cuts.  The approximate algorithm has running-time $O(n)$ for an $n$-individual pedigree on which to estimate the kinship coefficients of $\sqrt{n}$ individuals from $\sqrt{n}$ founder kinship coefficients.

\end{abstract}

\section{Introduction}

Computing the kinship coefficients is fundamental to understanding breeding relationships and human genealogies~\cite{thompson1985}.  The kinship coefficients have been used for correcting GWAS case-control studies for known relationships and for reducing spurious disease associations~\cite{Yu2005,Rakovski2009,Kang2010,Cordell2014}.  They have also been used for pedigree case-control disease association~\cite{Thornton2007}.  The legal permissibility of marriages can be defined using kinship coefficients that prevent inbreeding.  When breeding pedigree animals such as dogs, cats, horses, cows, or endangered species, using kinship coefficients to prevent inbreeding would foster the genetic diversity.  This genetic diversity is important, because it prevents excessive homozygosity and the raise of recessive Mendelian disease in the population as a whole~\cite{Woods2006}.

There is some confusion as to how quickly kinship coefficients can be computed~\cite{Abney2009}.  The kinship coefficients are defined by counting paths of genetic transmission of alleles.  While there exists a nice set of recursive equations, the algorithmic properties have needed more treatment in the literature.  This paper aims to fill this gap by introducing the three fastest known kinship algorithms: two exact algorithms and one approximate algorithm.

\section{Background}
\label{sec:background}

The pedigree graph, $P=(V,E)$, is the canonical descriptor of known relationships.  This is a directed acyclic graph with individuals as nodes, $V$, and edges, $E$, directed from parent to child.  The graph has in-degree at most two with one parent of each gender.  Let $I \subseteq V$ be the \emph{individuals of interest}, or the individuals whose relationships we would like to quantify.

There are a variety of ways to summarize the relationships in a pedigree in condensed forms.  One can look at a pair-wise pedigree relationships and describe those using the language of family relationships: parent, child, nephew, cousin, etc.  One can look at pair-wise genetic relationships and ask whether two alleles are inherited from an identical allele in an ancestor, i.e.~whether the alleles are identical-by-descent (IBD).  One can look at the $k$-wise relationships describing the probability of $2k$ alleles being IBD which is generalized kinship coefficient~\cite{Abney2009}.
One can also summarize a pair-wise relationship by the probability of two random alleles being IBD; this is the kinship coefficient, sometimes called the coefficient of relationship.  This last is the subject of this paper.

\begin{figure}[ht!]
  \begin{center}
  \includegraphics[width=3in]{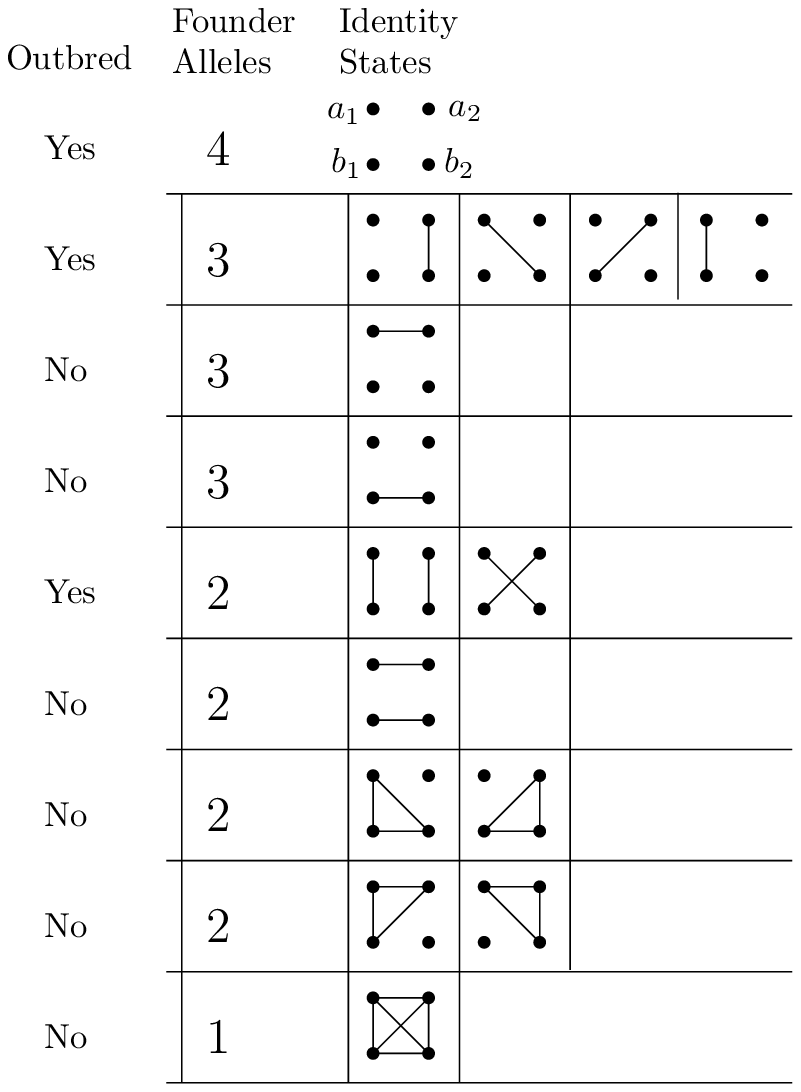}
  \end{center}
  \caption{{\bf Identity States.} Each identity state is a graph with four alleles of two individuals drawn as the nodes and edges appearing between every pair of IBD alleles.
The 15 identity states are grouped so that each row corresponds to one of the 9 condensed identity states.  The number of founder alleles for each identity state is listed, along with whether the identity state is out-bred.
}
  \label{fig:identstates}
\end{figure}

Formally, for two individuals in a pedigree, define \emph{identity-by-descent (IBD)} as the event that both individuals inherited an allele from the same ancestor.  For two individuals $i$ and $j$, there are four alleles, two for individual $i$, $a_1$ and $a_2$ and two for individual $j$, $b_1$ and $b_2$.  In complete generality, there are 15 ways for these 4 alleles to be IBD~\cite{Jacquard1972}.  We draw these 15 possibilities as graphs on 4 alleles, as in Figure~\ref{fig:identstates}.  There is an edge between every pair of alleles that is IBD.  For example, allele $a_1$ and $a_2$ can be IBD while none of the other alleles are IBD, see row 3 of Figure~\ref{fig:identstates}.  As another example, alleles $a_1$, $b_1$, and $b_2$ could be IBD while allele $a_2$ is not related to the others, see row 7 of Figure~\ref{fig:identstates} 

The \emph{kinship coefficient} of two individuals $i$ and $j$ in a a pedigree is the probability of identity-by-descent between two randomly drawn alleles, one from each person.  Let the matrix $\phi$ contain all the pair-wise kinship coefficients in a pedigree.  Entries $\phi_{ij}$, for $i \ne j$, are kinship coefficients, and entries $\phi_{ii}$ are related to inbreeding coefficients.  
Label the alleles of individuals $i$ and $j$ with distinct labels: $(a_1, a_2)$ and $(b_1,b_2)$, respectively.  
By the definition of kinship, 
\[ 
\phi_{ij} = (1/4)Pr[a_1 = b_1] + (1/4)Pr[a_1 = b_2] + (1/4)Pr[a_2 = b_1] + (1/4)Pr[a_2 = b_2] 
\]
where this $=$-operator means IBD.  Unlike the IBD probabilities which can apply to specific alleles in data, the kinship coefficients are an expectation over the structure of the pedigree and are independent of the data.
The kinship coefficients are often defined as an expectation over all possible inheritance paths (or all possible joint assignments to the segregation indicators of a pedigree), this paper will develop the equivalence with an expectation over the identity states and their probabilities.

Suppose that we consider only out-bred IBD possibilities (defined as having $\phi_{ii} = 1/2$ for all $i$).
The identity states encode all the possibilities for IBD, including inbreeding possibilities, and Figure~\ref{fig:identstates} has three rows indicating they are out-bred (rows 1,2,5).  These correspond to the a pair of individuals, $i$ and $j$ sharing zero, one, or two alleles IBD, respectively.  In this case, we can denote the probability of each of these event as $f^{ij}_0$, $f^{ij}_1$, and $f^{ij}_2$ respectively.  In this particular case, the kinship coefficients are simple to compute:
\[
  \phi_{ij} = (2f^{ij}_2 + f^{ij}_1) / 2 .
\]

Assume that there might be inbreeding in the pedigree.  Now, we wish to compute all the kinship coefficients in one computation.  The recursive algorithm for computing this was developed in detail in \emph{Pedigree Analysis in Human Genetics}~\cite{thompson1985} and in a 1981 paper by Karigl~\cite{Karigl81}.  This section will only give the recursive equations which are a top-down recursion on the pedigree.  For all founders $f$, the matrix is initialized as 
\begin{eqnarray*}
  \phi_{ff} &=& 1/2, \textrm{ and } \\
  \phi_{fj} &=& 0,  \textrm{ for any individual }j\textrm{ that is not a descendant of }f.
\end{eqnarray*}
Let the mother and father of $i$ be denoted $m$ and $p$.  Then use the kinship coefficient between $j$ and $i$'s parents to compute
\begin{eqnarray}
\label{eqn:kinship_from_parents}
  \phi_{ij} &=& (\phi_{mj} + \phi_{pj})/2 ~ \textrm{where} ~i~\textrm{is not an ancestor of}~ j ~\textrm{and}~ i \ne j
\end{eqnarray}
In a similar manner, we compute the kinship coefficient for $i$ from the $i$'s parents:
\begin{eqnarray}
  \phi_{ii} &=& (1 + \phi_{mp})/2
\end{eqnarray}
where this function of $\phi_{mp}$ accounts for the probability of picking the same allele when uniformly choosing two alleles from the same person.

Some researchers write the kinship matrix as given by the output of the above recursion.  Others transform the matrix, so that it contains the inbreeding coefficients for each individual instead of the kinship coefficients.  This transformation, applied only to the diagonal, is
\begin{eqnarray*}
  \Phi_{ij} &=& \phi_{ij}  \textrm{ for all } i \ne j \\
  \Phi_{ii} &=& 2 \phi_{ii} - 1
\end{eqnarray*}
We find it convenient to represent the inbreeding coefficient on the diagonal, and this is the convention used in this paper.

\section{Methods}

Before introducing the details of the algorithms, we need to develop some mathematical methods that ease the task of algorithm development.  First, the kinship recursion can be initialized several ways.  If the founders are known to be inbred, it is appropriate to initialize the recursion with that information, for more accurate computation of kinship coefficients.  Second, the kinship coefficients can be represented in terms of identity state coefficients.  This relationship is useful for developing a sampling algorithm that samples identity states.

\subsection{Generalized Initialization of the Kinship Recursion}
\label{sec:initialize}
We related the kinship to inbreeding coefficients estimated for unrelated individuals.  The founders of a pedigree have an ancestry that relates them, even if that ancestry is not recorded in the pedigree graph.  Knowing the kinship coefficients of the founders is the correct way to initialize the algorithm.

Let $\Psi$ be the kinship coefficients for the founders, i.e.~a matrix of $F \times F$ where $F$ is the number of founders.
\begin{eqnarray*}
  \phi_{ff} &=& (1+\Psi_{ff})/2,  \\
  \phi_{fg} &=& \Psi_{fg} \textrm{ for founders $f$ and $g$, $f \ne g$}, \textrm{ and } \\
  \phi_{fj} &=& 0,  \textrm{ for non-founder, not a founder child, }j\textrm{ that is not a descendant of }f.
\end{eqnarray*}
Recall that the diagonal elements of $\Psi$ are the inbreeding coefficients.
The recursive equations also need to be slightly modified, by the addition of the following case for founder children $c$ not descended from $f$
\begin{eqnarray*}
  \phi_{fc} &=& \phi_{cf} = (\phi_{m(c),f} + \phi_{f(c),g})/2 \textrm{ for founder $m(c)$ or $f(c)$ and founder $f$} .
\end{eqnarray*}

It my be difficult to obtain the kinship coefficients on the founders if their genealogy is unknown.  More feasibly, we can estimate the inbreeding coefficient from the homozygosity~\cite{Leutenegger2003}.  We modify the above recursion to initialize it with the average inbreeding among all the founders.  The initialization becomes
\begin{eqnarray*}
  \phi_{ff} &=& (1+\Psi_{ff})/2,  \\
  \phi_{fg} &=& \psi \textrm{ for founders $f$ and $g$, $f \ne g$}, \textrm{ and } \\
  \phi_{fj} &=& 0,  \textrm{ for non-founder, not a founder child, }j\textrm{ that is not a descendant of }f.
\end{eqnarray*}
where $\Psi_{ff}$ is the inbreeding coefficient for founder $f$, and $\psi = 1/F \sum_{h} \Psi_{hh}$ is the average inbreeding coefficient for founders $h$.  The change to the recursive equations is still relevant.

If the kinship coefficients of the founders are not properly taken into account, then the kinship will be computed assuming that the founders are out-bred.  Thus, it will inaccurately represent the genetic relationships between the individuals whose ancestry contains inbreeding in the founders.

\subsection{Relating Identity States and the Kinship Coefficient}

To relate the identity coefficients into kinship coefficients, we take an expectation over the identity states.  Since there is a deterministic mapping between the condensed identity states and the identity states, the expectation can be written in terms of the condensed identity states. 
The proof for this approach takes as a first step the expression of the kinship as an expectation over inheritance paths,~\cite{Kirkpatrick2011xxxx} which is easily converted into an expectation over identity states. % ISBRA
We will define this expectation, discuss how to compute it, and leave the proof as an exercise.

For an identity state, let $t \in \{aa, ab, bb\}$ denote an \emph{edge type}, for example, $ab$ indicates any edge between the alleles in two different individuals $a$ and $b$, i.e. and edge between one of the nodes $\{a_1, a_2\}$ and one of the nodes $\{b_1,b_2\}$. Another example, $aa$, indicates an edge between nodes $a_1$ and $a_2$ within the same individual.  Recalling that the identity state is a graph on nodes $\{a_1, a_2, b_1, b_2\}$, we let $e(s,t)$ be a function of identity state $s$ and edge type $t$ which gives the number of edges of type $t$ in identity state $s$. For example, in Figure~\ref{fig:identstates}, row 8, column 1, $e(s, aa) = 1, e(s, ab) = 2, e(s, bb) = 0$.

\newcommand\identitystates{{\mathcal S}}

The kinship coefficient $\Phi_{a,b}$ between individuals $a \ne b$ is related to the identity states and their coefficients via the following expectation over the set of 15 possible identity states (a set that we denote by $\identitystates$):
\begin{equation} \label{eq:phi_ab}
\Phi_{a,b} = \sum_{s\in\identitystates} \frac{e(s,ab)}{4} \; \bbP[S=s].
\end{equation}
In this equation, $e(s,ab) / 4$ can be interpreted as the fraction of the $4$ possible edges between one node in $\{a_1, a_2\}$ and one node in $\{b_1,b_2\}$ that are present in the identity state $s$.

The inbreeding coefficient $\Phi_{a,a}$ is computed slightly differently via:
\begin{equation} \label{eq:phi_aa}
\Phi_{a,a} = \sum_{s\in\identitystates} e(s,aa) \; \bbP[S=s],
\end{equation}
where $e(s,aa)$ indicates whether the single possible edge between nodes $a_1$ and $a_2$ exists.  

The transformation to obtain the kinship coefficient from Equation~\ref{eq:phi_aa} is
\begin{equation*} 
\phi_{a,a} = (1+\Phi_{a,a})/2 .
\end{equation*}
The convention in this paper is to use $\Phi_{a,a}$.

\section{Efficient Implementations}

We will discuss several algorithms for obtaining kinship coefficients which are efficient in different scenarios.  The first scenario is exact computation in polynomial time, and the second scenario is approximate computation in linear time.

First, the recursions given, above, can be implemented efficiently in an $O(n^2)$-time algorithm.  The challenging portion of the recursive equations is the check for which individuals are ancestors of each other.  This can be done in constant time, provided that the correct data structure tabulates the ancestry information.  The details of this algorithm are given in Section~\ref{sec:exact}.  

Second, when the individuals of interest are a subset of the individuals in the pedigree, then we can use a divide-and-conquer approach for applying the first exact algorithm.  In this approach we recursively cut the pedigree graph to obtain sub-pedigrees on which to apply the full exact algorithm.  As long as the individuals of interest all reside in a single sub-pedigree, this approach is more efficient than the first exact algorithm.  Details are given in Section~\ref{sec:recursivecut}

Third, when a linear algorithm is desired and the kinship coefficients of a subset of individuals are needed, there is a sampling algorithm that samples identity states.  The sampling algorithm works in $O(n)$ time provided that the kinship of $\sqrt{n}$ founders is given and that the kinship of $\sqrt{n}$ of the individuals is computed.  A smaller version of this sampling algorithm that estimates the kinship of one pair of individuals has previously been introduced~\cite{Sun2014}.  That algorithm was evaluated for sampling error, and it was discovered that several thousand identity states are needed for very large pedigrees.  Details of this algorithm appear in Section~\ref{sec:approximate}

\subsection{Efficient, Exact Algorithm}
\label{sec:exact}

There is an $O(n^2)$-time implementation of the kinship calculation where $n$ is the number of individuals in the pedigree, and $I = V$.  Note that any implementation that touches every cell of the kinship matrix requires running-time at least $O(n^2)$.  Any implementation of the kinship that represents the full kinship matrix requires $O(n^2)$ space.

We define $A_i$, the \emph{ancestor set} for individual $i$ in the pedigree, as the set containing $i$ and all its ancestors.  This object can easily be computed in $O(n^2)$ time.  The ancestor set allows us to do the ancestry check in the kinship recursion in constant time.

The ancestor sets are computed in a top-down recursion on the pedigree graph.  Initialize the recursion with 
\[
A_f = \{f\}
\]
where $f \in V$ is a founder in the pedigree, meaning the individual has no parents.

Now, the remaining individuals' ancestor sets are computed from their parents' as follows:
\[
A_i = \{i\} \cup A_{m(i)} \cup A_{f(i)}
\]
where $m(i)$ is the mother of $i$ and $f(i)$ is the father of $i$.

The top-down order is obtained by creating a queue of individuals, initializing the queue with the founders, and popping an individual from the front of the queue while simultaneously adding their children to the end of the queue.  This produces an order such that each individual is considered after all of their parents.

\begin{algorithm}
\caption{$O(n^2)$ Exact Kinship Algorithm}
\label{alg:exact}
\begin{algorithmic}
%\COMMENT{Initialize the ancestor sets.}
\STATE{Let $A$ be an $n \times n$ matrix initialized with all zeros.}
\FOR {all the founders $i \in V$}
  \STATE Let $A_{ii} = 1$
\ENDFOR
\FOR {each $i \in V$ with parents $m(i),f(i)$ such that their $A$ row is set}
  \FOR {each $j \in p(i)$}
    \FOR {each $v \in V$}
      \STATE $A_{iv} = A_{m(i)v} \textrm{ AND } A_{f(i)v}$
      \IF {$v == i$}
        \STATE $A_{ii} = 1$
      \ENDIF
    \ENDFOR
  \ENDFOR
\ENDFOR
%\COMMENT{Initialize the kinship recursion.}
\STATE Let $\Phi$ be an $n \times n$ matrix initialized with $-1$.
\FOR {every founder $f \in V$}
  \STATE $\Phi_{f,g} = (1+\Psi_{ff})/2$
\ENDFOR
\FOR {every pair of founders $f,g \in V$ with $f \ne g$}
  \STATE $\Phi_{f,g} = \Psi_{fg}$
\ENDFOR
\FOR {every founder $f$ and every non-founder $j \in V$, $j$ not founder child}
  \IF {$A_{jf} == 0$}
    \STATE $\Phi_{fj} = 0$
  \ENDIF
\ENDFOR
%\COMMENT{Run the kinship recursion, top-down.}
\FOR {every $i \in V$ whose parents have been assigned kinship}
  \FOR {every $j \in V$ whose parents have been assigned kinship}
    \IF {$i == j$}
      \STATE $\Phi_{ii} = (1+\Phi_{mp})/2$
    \ELSE
      \IF {$A_{ji} == 0$}
        \STATE $\Phi_{ij} = (\Phi_{mj} + \Phi_{pj})/2$
      \ENDIF
    \ENDIF
  \ENDFOR
\ENDFOR
\FOR {every $i \in V$}
  \STATE $\Phi_{ii} = (2*\Phi_{ii})-1$
\ENDFOR
\end{algorithmic}
\end{algorithm}

The recursion from Section~\ref{sec:background} can now be implemented in $O(n^2)$ time, using the ancestor sets.  The ancestor sets can be queried quickly, if they are stored in a look-up table.  This makes the recursion a double loop over the individuals in the pedigree, when those individuals are considered in the top-down order.  Algorithm~\ref{alg:exact} gives the details of the recursive loops.

\subsection{Faster, Recursive-Cut Exact Algorithm}
\label{sec:recursivecut}

In the case where we require the kinship coefficients of a subset of the pedigree individuals, $I \subset V$, we can improve the running-time for the exact calculation.  Recall from Section~\ref{sec:initialize}, that a correct kinship computation can be done after initializing the founders of a pedigree with their known kinship coefficients.  Consider a large pedigree, which is a directed acyclic graph with the founders as sources and the leafs as sinks.  Informally, we can segment the large graph by taking cuts that run 'vertically' across a generation.  The individuals in the cut become the leafs of the upper segment and the founders of the lower segment of the pedigree.  This splits the pedigree in half at the cut, and allows us to apply the exact algorithm from Section~\ref{sec:exact} to each segment of the pedigree.

Formally, a \emph{graph cut} is defined as a partition of the vertices of a graph into two sets such that the \emph{cut edges}, when removed, disrupt every path from any source to any sink.  A graph cut produces two disjoint subgraphs.  In our application we will need to slightly modify the sub-pedigrees so that they are no longer disjoint. 

A \emph{pedigree cut} for the purposes of this algorithm will be defined as a a set of cut edges and the set of cut edges defines a generation of individuals.  Any pedigree cut separates the pedigree into two sub-pedigrees, one containing the founders which we will refer to as the upper, or older, sub-pedigree, and one containing the leafs which we will refer to as the lower, or younger, sub-pedigree.  The upper sub-pedigree, by the placement of the pedigree cut, now has new leaf individuals, which are the parent nodes of the cut edges.  Let the parents be set $P \subseteq V$.  For our application, the lower sub-pedigree needs to contain set $P$ as the founders.  Therefore the lower sub-pedigree contains the set of cut edges.

We can recursively bipartition the pedigree and sub-pedigrees many times to get a reasonable running time of the exact kinship algorithm on any single sub-pedigree.  The finest recursive partitioning will produce generational sub-pedigrees.  The number of generations puts a lower bound on the number of recursive cuts that are possible.  Also, the individuals of interest all need to appear in a single sub-pedigree, if we are to obtain all their pair-wise kinship coefficients from this algorithm.

Now, we apply the exact kinship algorithm the the sub-pedigrees from the top down.  The kinship coefficients for the leafs of each sub-pedigree are used to initialize the founder kinship coefficients for the next sub-pedigree.  This is done iteratively down the pedigree, until the kinship coefficients of the individuals of interest are obtained.

When applied to a generational pedigree drawn using the diploid Wright-Fisher model, the pedigree has $2N$ individuals per generation and $G$ generations.  This recursive-cut kinship algorithm runs on this generational pedigree in $O(N^2G)$ time.  More generally if we cut an arbitrary pedigree into $m$ segments with the maximum segment having $s$ individuals, then the running-time of the recursive-cut kinship algorithm is $O(s^2m)$.

The main disadvantage of both these exact algorithms is that they are only polynomial in running time and perhaps may not be fast enough for applications on very large pedigrees.  For very large pedigrees, we need a linear-time algorithm for scalability.

\subsection{Fast, Approximate Algorithm}
\label{sec:approximate}
There is an $O(n)$-time algorithm for $I \subset V$ with $|I| = O(\sqrt{n})$ and $F = O(\sqrt{n})$ for number of founders $F$, and this algorithm quickly estimates the kinship coefficients by sampling identity states and using the expectations in Equations~\ref{eq:phi_ab} and~\ref{eq:phi_aa}.

Recall that in diploid individuals, each individual has two alleles at every site in the genome.  Of these two alleles, one comes from the father, and one from the mother.  From each parent, the allele is copied either from the grand-father or from the grand-mother.  This binary choice of grand-paternal origin is usually stored in \emph{segregation} indicators, $x_i = (x^m_i,x^f_i)$, for individual $i$ where $x^p_i \in \{m,f\}$.  An inheritance path consists of the segregation indicators of all the individuals in the pedigree.

Consider the graph of all the alleles of all the individuals at one site with edges connecting the alleles that are inherited from parent to child (i.e.,~the edges indicated by the segregation indicators).  This is a graph of the inheritance path, and it has connected components.  Let the \emph{CC membership} of each allele be given by a tuple of integers, $c_i = (c^m_i,c^f_i)$, for individual $i$.

\begin{algorithm}
\caption{$O(n)$ Approximate Kinship Algorithm for $|V|=n$, $|I| = O(\sqrt{n})$, and $F = O(\sqrt{n})$}
\label{alg:approximate}
\begin{algorithmic}
%\COMMENT{Sample the inheritance path}
\STATE Let $\Phi$ be the $n \times n$ matrix initialized with zeros.
\FOR {$s$ in 1 to $S$ where $S$ is number of samples}
\FOR {$i \in V$}
  \STATE Flip two $\{m,f\}$-labeled coins, and assign their values to $(x^m_i,x^f_i)$.
\ENDFOR
%\COMMENT{Get the CC membership}
\STATE Initialize all $c_i = (0,0)$
\STATE Let $counter = 1$.
\FOR {$l \in V$ where $l$ is a leaf}
  \STATE Let $(c^m_i,c^f_i) = (counter, counter+1)$
  \STATE $counter = counter + 2$
\ENDFOR
\FOR {each $i \in V$ provided that all of $i$'s children have been processed}
  \FOR {$p \in \{m,f\}$ and $j \in V$ being the appropriate gendered parent}
     \IF {$c^{x^p_i}_j < c^p_i$}
       \STATE $c^{x^p_i}_j = c^p_i$
     \ENDIF
  \ENDFOR
\ENDFOR
\FOR {every pair of founders $f$ and $g$}
  \STATE flip a coin for edges $(a_1,b_1)(a_2,b_2)$ or $(a_1,b_2)(a_2,b_1)$, $a=0$ or $a=1$, respectively
  \FOR {$x \in \{m,f\}$, and let $y \in \{m,f\} \setminus x$}
    \STATE Let $u \in [0,1]$ be a uniform random variable
    \IF {$u \le \Psi_{fg}$}
      \IF {$a==0$}
        \STATE $c^x_f = c^x_g$
      \ELSE
        \STATE $c^x_f = c^y_g$
      \ENDIF
    \ENDIF
  \ENDFOR
\ENDFOR
\FOR {each $i \in V$ provided that all $i$'s parents have been processed}
   \STATE $c^p_i = c^{x^m_i}_p$
\ENDFOR
%\COMMENT{Do kinship updates}
\FOR{each $i \in I$}
  \FOR{each $j \in I$}
    \STATE Create the identity state graph for $(c_i,c_j) = (c^m_i,c^f_i,c^m_j,c^f_j)$
    \IF {$i == j$}
      \IF{$c^m_i == c^f_i$}
        \STATE $\Phi_{ii} = \Phi_{ii} + \frac{1}{|S|}$
      \ENDIF
    \ELSE
        \STATE Let $e$ be the number edges between $c_i$ and $c_j$ in the identity state graph.
        \STATE $\Phi_{ij} = \Phi_{ij} +  \frac{e}{4*|S|}$
    \ENDIF
  \ENDFOR
\ENDFOR
\ENDFOR % for each sample
\end{algorithmic}
\end{algorithm}

Our method, Algorithm~\ref{alg:approximate}, will sample an inheritance path with one pass through the pedigree.  When considering each individual, the algorithm flips a coin to set the segregation indicators for that individual.  Later, these indicators will be used to determine which alleles are identical-by-descent for this inheritance path.

Two more passes through the pedigree will be used to set the CC membership with consistent values.  The leaf alleles of the pedigree are populated with distinct integers representing their putative CC membership.  The first pass proceeds from the bottom of the pedigree to the top, and processes each child before their parents.  For each allele in each individual, their CC membership integer is copied to the allelic ancestor given by the segregation indicator.  If two children give two different CC memberships to the parent, the tie is broken with the largest integer value.  To account for founder inbreeding, we merge founder CC memberships randomly according to the probabilities given by the initialization kinship coefficients.  The second pass is down the pedigree from top to bottom, and the tie-broken CC membership integers are simply copied back down the path of allelic inheritance.  After these two passes, every allele in a connected component of the inheritance path graph will have the same CC membership, and every pair of alleles from different CC's will have distinct CC membership unless the CC's were randomly merged by founder inbreeding.

Finally, for every pair of individuals of interest, we can use the CC membership to obtain the identity state.  This identity state can be used to update the estimated kinship with the appropriate terms from Equations~\ref{eq:phi_ab} and~\ref{eq:phi_aa}.

The details of these procedures are left to Algorithm~\ref{alg:approximate}.  The convergence properties of this sampling algorithm were explored using simulations by Sun, et al~\cite{Sun2014}.  For very large pedigrees, they found that thousands of samples might be required.

\section{Discussion}

The above kinship algorithms are applicable to large pedigrees, since they either have efficient polynomial or linear running-times.  This paper gives two exact algorithms and one approximation algorithm.  The fastest known exact algorithm for kinship computations is the recursive-cut exact algorithm.  All these algorithms are easily run using the source code that is published in tandem with this paper.

All of the algorithms in this paper have running times that are parameterized by the number of individuals.  Since the kinship coefficients are defined by inheritance paths which correspond to edges in the pedigree graph, it is possible that the number of edges in a pedigree graph provides the true lower-bound on the number of operations needed to compute the exact kinship coefficients.  We leave it as an open problem whether there is an efficient algorithm, perhaps along the lines of the recursive-cut algorithm, that has running-time parameterized by the number of edges in the pedigree graph.

Another open problem is whether there is a sparse algorithm for computing the kinship.  This type of algorithm would only compute the non-zero entries in the kinship matrix.  If designed properly, such an algorithm would need far less space, since it would not need to represent the entire kinship matrix.  We believe that such an algorithm does exist and has yet to be discovered.

\section{Code}

The above polynomial-time exact algorithm and the linear-time approximation algorithms are implemented as PedKin in C++ and are available under the GNU GPL v2.0 open source license.  
This document is to be distributed with the source code.
The PedKin source code is available at: \\
{\tt http://www.intrepidnetcomputing.com/research/code/}.

\bibliography{kinbib}
\bibliographystyle{plain}

 \end{document}